\documentclass[11pt]{article}
\usepackage{moriond,epsfig}

\def\be{\begin{equation}}
\def\ee{\end{equation}}
\def\bea{\begin{eqnarray}}
\def\eea{\end{eqnarray}}

\begin{document}
\vspace*{4cm}
\title{Neutrino Astronomy beyond and beneath the Horizons}

\author{D.Fargion}

\address{Department of Physics and INFN, Rome University La Sapienza,Pl.A.Moro 2,\\
Rome 00185, Italy}

\maketitle\abstracts{ The Earth's Shadow to Cosmic Rays offer a
windows to Tau Neutrino Astronomy at the Horizon edges. Inclined
and Horizontal C.R. Showers ($70^o-90^o$ zenith angle) produce
secondary ($\gamma$,$e ^\pm$) mostly suppressed
 by high column  atmosphere  depth. The shower  Cherenkov  photons
 are diluted and filtered by air opacity,  but secondary penetrating,
  $\mu^\pm$  and  decay into $e^\pm$,$\gamma$, revive additional Cerenkov flash lights. The
larger horizontal distances widen the shower's cone while the
geo-magnetic field open it in a  very characteristic fan-like
shape polarized by local field vector. These elongated showers
jets are more frequent, up to $10^3$ (respect to vertical
showers). Most recent and largest GeVs $\gamma$ telescopes   may
detect such UHECR horizontal showers at energies PeVs up to EeV
(or higher) . Details on arrival angle and column depth, shower
shape, timing signature of photon flash intensity, may inform  on
the altitude interaction and on primary UHECR composition. At
larger zenith angle ($90^o-99^o$) among single albedo muons, more
rare up-going showers are traced by muon (as well as
$e^\pm$,$\gamma$) bundles that would give evidence of rare
Earth-Skimming neutrinos, ${\nu}_{\tau}$, $\overline{\nu}_{\tau}$,
at PeVs-EeVs energies. They are arising by Tau Air-Showers
(HorTaus) (${\nu}_{\tau} + N \rightarrow \tau + X $,
$\tau\rightarrow $ hadrons and/or electromagnetic shower far from
their Earth exit). Their rate may be comparable with $6.3$ PeVs
$\overline{\nu}_{e}-e$ neutrino induced air-shower originated
above and below horizons, in interposed atmosphere, by $W^-$
resonance at Glashow peak. Additional and complementary UHE SUSY
$\chi^o + e\rightarrow \tilde{e} \rightarrow\chi^o + e $ at tens
PeVs-EeV energy may blaze, as $\overline{\nu}_{e}-e \rightarrow
W^-$ shower. Also UHE neutrino interacting in matter may lead to
$\tilde{\tau}$ whose escape from Earth might be source of
$\tilde{\tau}$-Air-showers. TeV new gravity interactions will
amplify these events. Their complementary nature might be
disentangled by Stereoscopic Telescopes  array experiment.
Surprisingly  Magic and Hess telescope in  present set up may be
already comparable to AMANDA underground neutrino detector at PeVs
energies looking at night to the Earth edges; such telescopes
pointing at the horizons toward active sources (AGN,BL Lac
Objects, GRBs or SGRs blazing micro-quasar jet, and SNRs) it does
experience in those directions an air mass corresponding to a
$km^3$ water one. Therefore Horizontal Showering is already the
most sensitive windows to Neutrino Astronomy.}

\section{UHECR traced by Cerenkov Lights  and Muon Bundles at Horizons}
Ultra High Energy Cosmic Rays (UHECR) Showers (from PeVs up to
EeVs and above, mainly of hadronic nature) born at the high
altitude in the atmosphere, may blaze (from the far edge)
\emph{above  the horizon} toward Telescopes such as Magic one. The
earliest gamma and Cerenkov lights produced while they propagate
through the atmosphere are severely absorbed because of the deep
horizontal atmosphere column depth ($10^4 - 5 \cdot 10^4$ $ g
\cdot cm^{-2}$ ),
  must anyway survive and also revive: indeed additional diluted but  penetrating
   muon bundles (from the same by C.R. shower)
   are decaying not far from the Telescope into electrons which are source themselves of small Cerenkov lights .
  Direct muons hitting the Telescope may blaze a ring
  or an arc of  lights.  These  muon bundle secondaries, about $10^{-3}$ times
  less abundant than the  peak of the gamma shower photons are  arising at high altitude, at an horizontal
  distances  $100-500$ km far from the observer (for a zenith angle
  $85^o-91.5^o$ while at 2.2km. height);
  therefore their hard (tens-hundred GeV) muon shower bundles (from  ten to millions muons at TeVs-EeVs C.R. energy primary)
   might spread in huge areas (up to tens- hundred $km^2$); they are  marginally bent by geo-magnetic fields
    and they are randomly scattered, often decaying at hundred-tens GeV energies,
     into electrons pairs; their consequent mini electromagnetic-showers are traced by their optical Cerenkov flashes.
  These diluted (but spread and therefore better detectable) brief (nanosecond-microsecond)
   optical signals may be captured as a light cluster by largest  telescope on ground
    as recent Stereoscopic Magic,  Hess, Veritas arrays. Cerenkov flashes, single or  clustered,
   must take place, at detection threshold, at least at a rate of  hundreds events a night
   for Magic-like Telescope facing horizons at zenith angle  $ 85^o \leq \theta \leq 90^o$.
   Their "guaranteed" discover may offer  a new tool in  CR and UHECR detection.
   Their primary hadronic signature might be hidden by the deep column depth distance,
    but there is a new trace offered by
   its secondary muon-electron-Cerenkov flashes in flight. On the same time  \emph{below the horizons}  a more rare (three-four order of
 magnitude)  but  more exciting PeV-EeVs Neutrino ${\nu_{\tau}}$
   Astronomy may arise by the Earth-Skimming Horizontal Tau Air-Showers (HorTaus); these UHE Taus are produced
   inside the Earth Crust by the primary UHE incoming neutrino
      ${\nu}_{\tau}$, $\overline{\nu}_{\tau}$,
       generated mainly by their muon-tau neutrino oscillations from
       galactic or cosmic sources,\cite{Fargion 2002a},\cite{Fargion03},\cite{Fargion2004}.
   Above or below the horizon edge, within a
   few hundred of km distances, horizontal showers could reveal the
   guaranteed  tuned   $\overline{\nu}_e$-$e\rightarrow W^-\rightarrow X$ air-showers at $6.3 PeV$ Glashow resonant peak
    energy; the $W^-$ main  hadronic ($2/3$) or leptonic and electromagnetic ($1/3$) signatures
    may be well observed (within tens-hundred km distance)  and
     their rate might calibrate a new horizontal neutrino-multi-flavour
     Astronomy \cite{Fargion 2002a}. The  $\overline{\nu}_e$-$e\rightarrow W^-\rightarrow X$
     of nearby nature (respect to most far away ones at same zenith angle of hadronic nature) would be better revealed by
     a Stereoscopic Magic  twin telescope or a Telescope array like Hess, Veritas.
       Additional Horizontal flashes  might arise
    by Cosmic UHE $\chi_o + e \rightarrow  \widetilde{e}\rightarrow \chi_o +
    e$ electromagnetic showers  within most SUSY models, if UHECR are born in topological
   defect decay or in their annihilation, ejecting a relevant component of SUSY
   particles. The UHE $\chi_o + e \rightarrow  \widetilde{e}\rightarrow \chi_o +
    e$ behaves (for light $\widetilde{e}$ masses around Z boson ones)
    as the Glashow  resonant case \cite{Datta}.
    Finally similar signals might be abundantly and better observed if UHE neutrinos share new extra-dimension
    (TeV gravity) interactions: in this case also  neutrino-nucleons interaction may be an abundant source of PeVs-EeVs
    Horizontal Showers originated in Air \cite{Fargion 2002a}. The total amount of air
    inspected within the solid angle $2^o \cdot 2^o$ by MAGIC height  at  Horizons ($360$ km.) exceed
    $44 km^3$ but their consequent detectable beamed volume  are  corresponding
    to an isotropic  narrower volume:  V$ = 1.36 \cdot 10^{-2}$ $km^3$ , nevertheless comparable
     (for Pevs $\overline{\nu}_e$-$e\rightarrow W^-\rightarrow X$
      and EeVs  ${\nu}_{\tau}$, $\overline{\nu}_{\tau} + N \rightarrow \tau\rightarrow $ showers)
       to the  present AMANDA confident volume. Moreover
       monitoring a defined source (like Crab,AGN,BL Lac, GRBs,
       SGRs, possibly a  micro-quasar jet and SNRs) at its dawn or rise,
       the horizontal interposed conical air volume $V_{air} \simeq 1000 km^3$
       exceed the $km^3$ water mass, making such Horizontal
       Showering Astronomy (by Magic-like telescopes) already
       the most sensitive Neutrino detector .

\begin{figure}
\begin{center}
\epsfig{figure=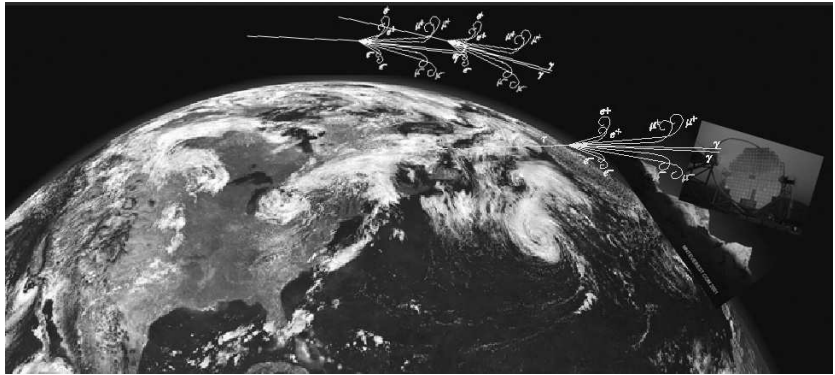,width=11cm}
\end{center}
\caption{ Schematic Picture of an Horizontal Cosmic Ray Air-Shower
(superior track) (HAS),  and an up-going Tau Air-Shower induced by
EeV Earth-Skimming $\bar{\nu_{\tau}}$,$\nu_{\tau}$ HORTAU and
their muons and Cerenkov lights blazing a Telescope as the Magic
one. Also UHE $\bar{\nu_e}-e$ and $\chi^o - e$ Scattering in
terrestrial horizontal atmosphere at tens PeVs energy may simulate
HAS, but mostly at nearer distances respect largest EeV ones of
hadron nature at horizon's edges. }
\end{figure}

\begin{figure}
\begin{center}
\epsfig{figure=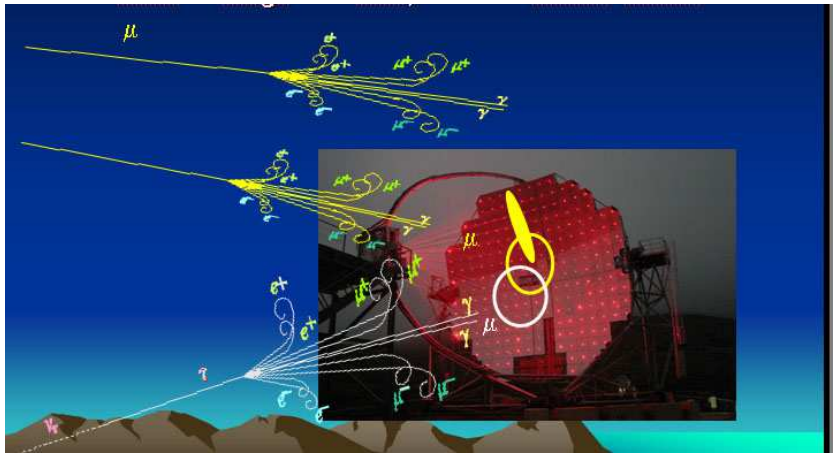,width=11cm}
\end{center}
\caption{ Schematic Picture as above for direct muon and lateral
gamma-electron showers by secondary decaying muons }
\end{figure}

\begin{figure}
\begin{center}
\epsfig{figure=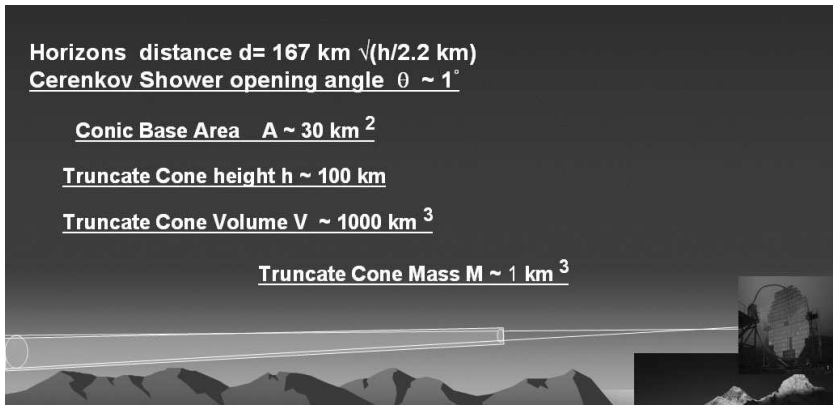,width=11cm}
\end{center}
\caption{ Schematic Picture of an Horizontal Cosmic Ray Air-Shower
( induced by  UHE $\bar{\nu_e}-e$ (at resonant energy $6.4$ PeV)
and $\chi^o - e$ scattering in terrestrial horizontal atmosphere
at tens PeVs energy. The distances and consequent  volumes within
the view cone exceed the $10^3$ $km^3$ air volume and a mass
comparable with a  $km^3$ water or ice mass. Therefore MAGIC ,
while  pointing a GRB or a SGR  Burst at Horizons ($~  3\% $ ) of
the GRB-SGRs events  behave as a km3 neutrino telescope. }
\end{figure}

\begin{figure}
\begin{center}
\epsfig{figure=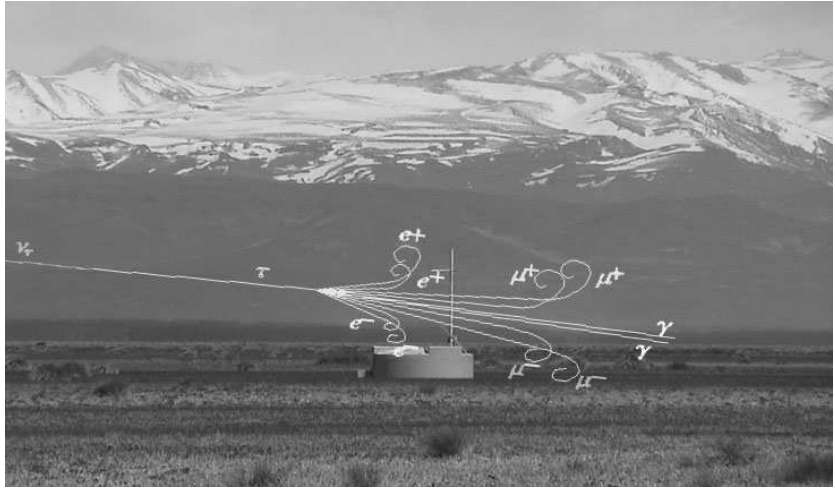,width=11cm}
\end{center}
\caption{ Schematic Picture of an Horizontal Cosmic Ray Air-Shower
(superior track) (HAS),   induced by EeV
$\bar{\nu_{\tau}}$,$\nu_{\tau}$ HORTAU  born inside the east side
of the Ande at west side of Auger array detector. Our prediction
is that within the first year of record AUGER detectors $must$
reveal the Ande shadows to UHECR, while within three years it
$might$ observe nearly two decades of HORTAUs induced by GZK
neutrinos inside the same mountain chain (Fargion et. all
1999.2004)}
\end{figure}

\begin{figure}
\begin{center}
\epsfig{figure=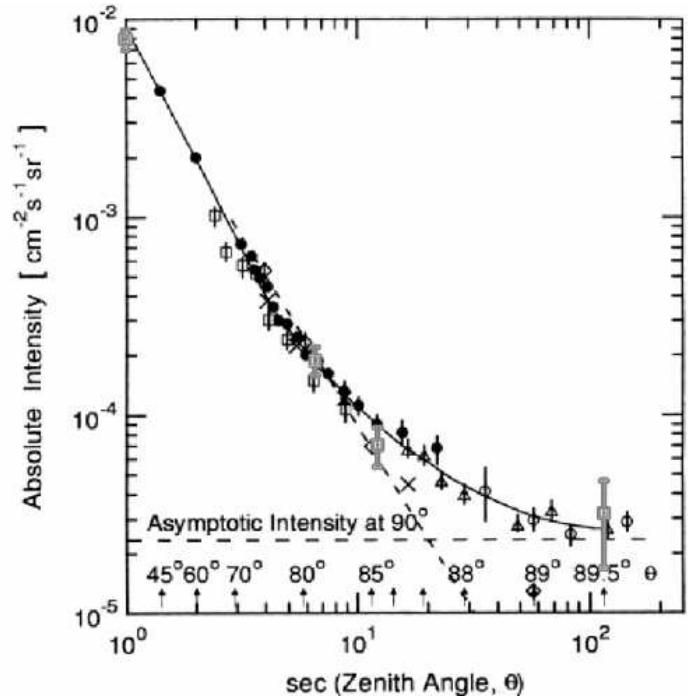,width=10cm}

\end{center}
\caption{ Observed Flux of Muons as a function of the zenith angle
above (see Iori ,Sergi,Fargion 2004 and,Grieder 2001)  the
horizons; for the muons below the horizons their flux at $91^o$
zenith angle is two order of magnitude below $\simeq 10^{-7}
cm^{-2} s^{-1} sr^{-1}$, as observed by NEVOD and Decor detectors
in recent years. As the zenith angle increases the upward muons
flux reduces further ; at $94^o$ and ten GeV energy it is just
four order below: $\simeq 10^{-9} cm^{-2} s^{-1} sr^{-1}$, see
recent results by NEVOD and DECOR experiments (2003) ; at higher
energies (hundred GeVs) and larger zenith angle only muons induced
by atmospheric neutrinos arises at $\simeq 2-3\cdot 10^{-13}
cm^{-2} s^{-1} sr^{-1}$ as well as Neutrino Tau induced Air-Shower
(muon secondaries)}.
\end{figure}

\begin{figure}
\begin{center}
\epsfig{figure=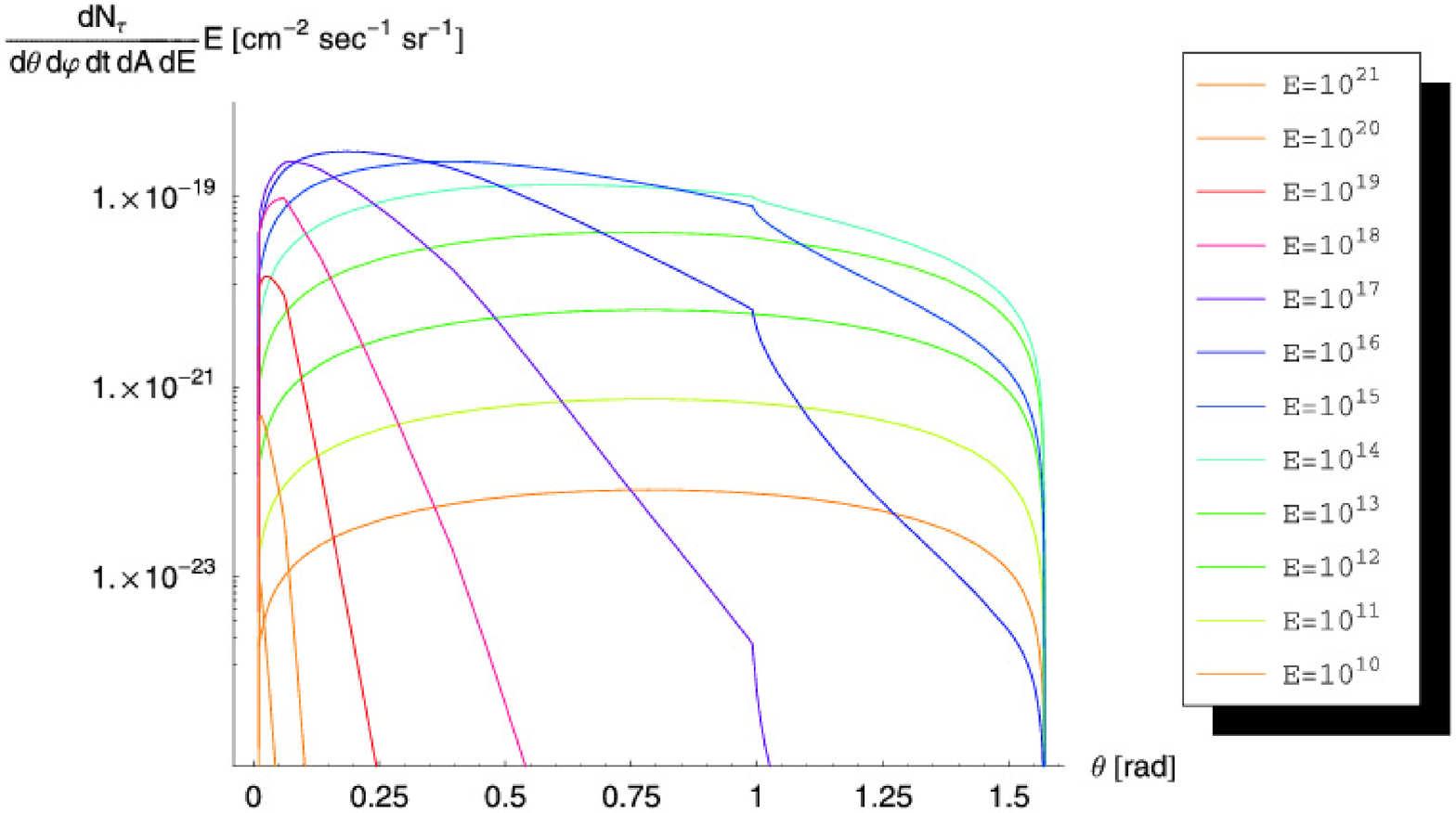,width=12cm} \caption{ Tau
Air-Showers  rates angular distribution, by Earth Skimming
Neutrino $\tau$ inside the rock, assuming a rock surface density;
the UHE $E_{\nu_{\tau}}$  energies are shown in eV unity. The
incoming neutrino flux  is a minimal GZK flux  (for combined
$\nu_{\tau}$ and $\bar{\nu_{\tau}}$ ) at $\phi_{\nu{\tau}}\cdot
E_{\nu_{\tau}} \simeq $ $50$ eV $cm^{-2}s^{-1} sr^{-1}$.
}
   \label{fig3}
\

\epsfig{figure=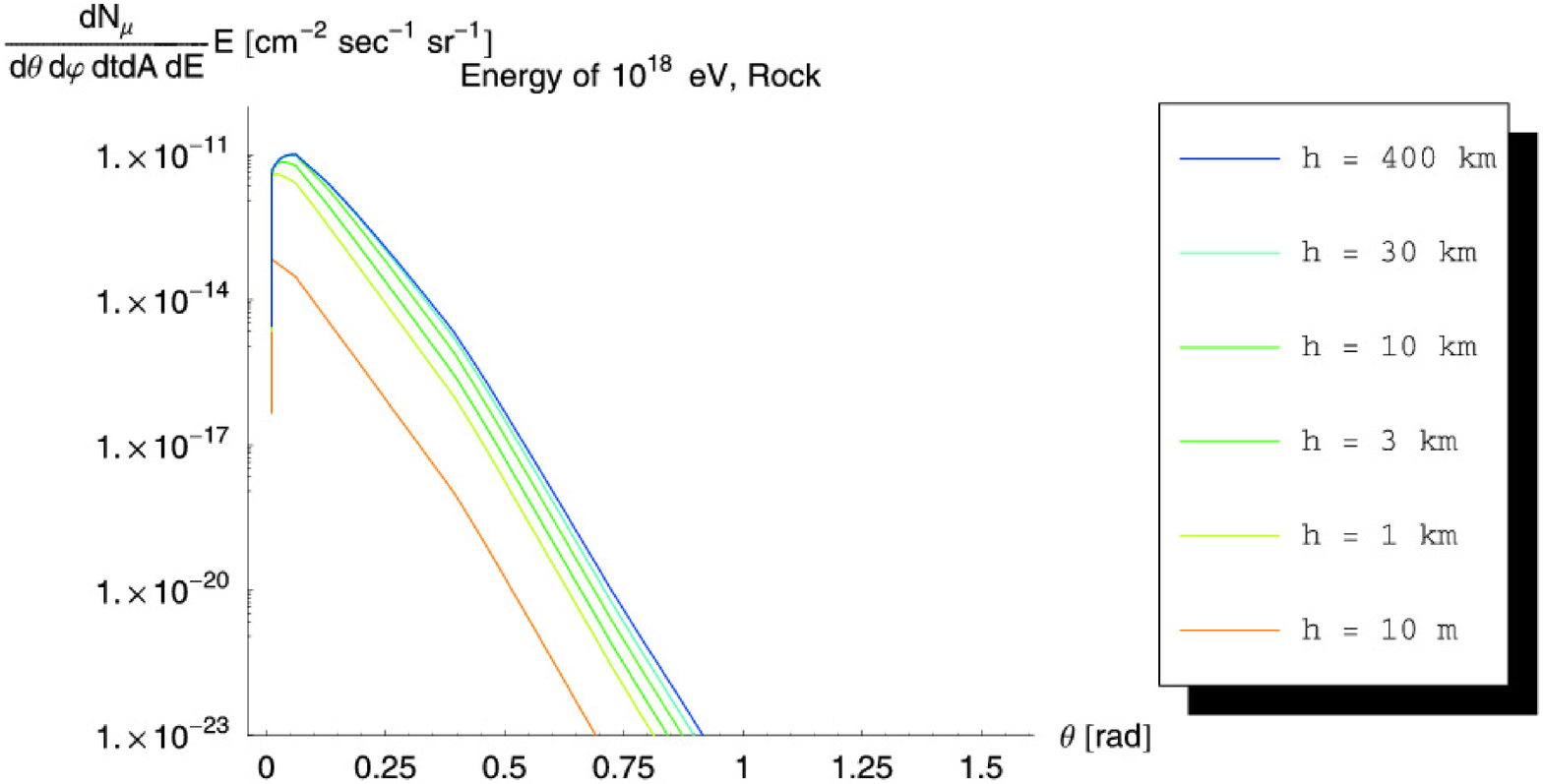,width=12cm} \caption{ Consequent
 Muons Secondary (by Tau Air-Showers) rate angular distribution at different
observer quota height (see label), at $10^{18}$eV energy,
exceeding (at horizontal zenith angle $\theta \simeq 93-97 ^o$ )
even the same atmospheric neutrino induced up-going muon  flux
$\phi_{\mu} \simeq 3 \cdot  10^{-13} \cdot cm^{-2} s^{-1}sr^{-1}$.
The incoming neutrino flux  is a minimal GZK flux (for combined
$\nu_{\tau}$ and $\bar{\nu_{\tau}}$ ) at $\phi_{\nu{\tau}}\cdot
E_{\nu_{\tau}} \simeq $ $50$ eV $cm^{-2}s^{-1} sr^{-1}$;
}
   \label{fig4}
   \end{center}
\end{figure}

\section{ Blazing Cerenkov Flashes by Horizons Showers and decaying Muons }
  The ultrahigh energy cosmic rays (UHECR) have been studied
   in the past mainly versus their secondaries ($\gamma$, $e^\pm$, $\mu^\pm$)
  collected vertically in large  array detectors on the ground. This is due to the rare event rate
  of the UHECR  in the atmosphere and due to the high
  altitude where the shower takes place, expand and amplify downward.
  On the contrary at the horizons the UHECR  are hardly observable (but also rarely looked
  for).  They are diluted both by  the larger  distances as well as by the exponential
  atmosphere opacity suppressing the electromagnetic (electron pairs and gamma) secondaries;
   also their rich optical Cerenkov signal is partially  suppressed by the horizontal air opacity.
   However this suppression acts also as an useful filter
  leading to the  higher CR events; their Cerenkov lights
  will be scatter and partially transmitted (around $90^o$ zenith angle by a factor $1.8\cdot 10^{-2}$ at $551$ nm., $6.6\cdot 10^{-4}$ at $445$ nm.)
  depending on the exact zenith angle and seeing: assuming  on average a suppression $5\cdot 10^{-3}$
  and  the  nominal Magic energy threshold at $30$ GeV , it  does
    corresponds to a hadronic shower at far horizons (diluted by nearly three order of magnitude by larger distances)
     at  energy above $E_{CR}\simeq 6 $ PeV.  Their primary flux may be estimated considering
     the known cosmic ray  on the top of the atmosphere (both protons
    or helium) (see DICE Experiment referred in\cite{Grieder01}) :
     $\phi_{CR}(E_{CR} = 6\cdot 10^{15} eV)\simeq 9\cdot
     10^{-12}cm^{-2}s^{-1}$.
  Within a Shower Cerenkov angle $\Delta\theta = 1^o $
     at a distance  $d =167 km \cdot \sqrt{\frac{h}{2.2 km}}$
     (zenith angle $\theta \simeq 87^o- 88^o$) the shower surface
     corresponds to a wide  area :
     $ [A = \pi \cdot(\Delta\theta \cdot d)^2\simeq 2.7 \cdot 10^{11} cm^2 (\frac{d}{167 km})^2 ]$, observed
     within an opening angle $[\Delta\Omega =(2^o \cdot 2^o)\pi \simeq 3.82  \cdot 10^{-3} sr.]$;
      the consequent event rate time  a night of record ($[\Delta(t)= 4.32 \cdot 10^4
      s]$) by Magic is    $$N_{ev}=\phi_{CR}(E= 6\cdot 10^{15} eV)\cdot A \cdot \Delta \Omega \cdot
      \Delta(t) \simeq 401/12 h$$. Therefore one may foresee  nearly every
      two minutes  a far hadronic Cerenkov lightening  Shower in Magic facing at the far horizons
      at zenith angle $87^o-88^o$.  Increasing the observer altitude h,
      the allowable    horizon zenith angle also grows: $\theta \simeq [90^o + 1.5^o \sqrt{\frac{h}{2.2km}}]$
       In analogy at a more distant horizontal edges (standing at height $2.2
       km$ as for Magic, while observing at zenith angle $\theta \simeq 89^o- 91^o$
         still above the horizons) the observation range $d$ increases : $d= 167\sqrt{\frac{h}{2.2 km}} + 360 km = 527
       km$;  the consequent shower area widen by more than an order of
       magnitude (and more than  three order respect to vertical showers) and the consequent foreseen event number,
       now for a much harder penetrating C.R. shower at $E_{CR} \geq 3\cdot 10^{17} eV$,  becomes:
        $$N_{ev}=\phi_{CR}(E= 3\cdot 10^{17} eV)\cdot A \cdot \Delta \Omega \cdot
      \Delta(t) \simeq 1.6 /12 h$$
       Therefore at the far edges of  the horizons $\theta \simeq 91.5^o$, once a night, an UHECR around EeV energies,
      may blaze to the Magic (or Hess,Veritas, telescope arrays).
        At each of these far primary Cherenkov flash is associated a
        long tail of secondary muons in a very huge area; these muons eventually are also hitting inside the
        Telescope disk; their nearby showering in air, while decaying
        into electrons in flight, (source of electromagnetic mini-gamma showers of  tens-hundred GeVs energy)
        is  also detectable at a rate discussed below.

\section{Single-Multi muon signals: Arcs, Rings and $\gamma$ Showers by
$\mu^\pm \rightarrow  e^\pm$}

    As already noted the main shower blazing photons from a CR  may be also regenerated  or aided by its secondary
    tens-hundred GeVs muons, either  decaying in flight as a gamma
    flashes,  or  directly  painting Cerenkov  arcs or rings while hitting the telescope.
    Indeed these secondary  penetrating muon bundles
    may reach hundreds km  distances ($\simeq 600 km \cdot\frac{E_{\mu}}{100\cdot GeV}$) far away from the shower origin.
    To be more precise a part of the muon primary energy will dissipate along $360$ km air-flight (nearly a
      hundred GeV energy), but a primary $130-150$ GeV  muon will survive at final
       $E_{\mu} \simeq 30-50$ GeV energy,  at minimal  Magic threshold value.
    Let us remind the characteristic secondary abundance in a shower:
    $ N_\mu \simeq 3\cdot 10^5 \left( \frac{E_{CR}}{PeV}\right)^{0.85}
    $.
     These secondaries  are mostly at a minimal (GeV) energies
\cite{Cronin2004};  for the harder (a hundred
   GeV) muons their number is (almost inversely proportionally to energy) reduced:
  $ N_\mu(10^2\cdot GeV) \simeq 1.3\cdot 10^4 \left( \frac{E_{CR}}{6 \cdot PeV}\right)^{0.85}$
   These values must be compared with the larger peak multiplicity (but much lower energy) of
   electro-magnetic shower  nature: $ N_{e^+ e^-} \simeq 2\cdot 10^7 \left(
   \frac{E_{CR}}{PeV}\right); N_{\gamma} \simeq  10^8 \left(
   \frac{E_{CR}}{PeV}\right) $.  As mentioned most of these electromagnetic tail  is lost (exponentially) at
  horizons (above slant depth of a few hundreds of
  $\frac{g}{cm^2}$)(out of the case of re-born, upgoing $\tau$ air-showers
  \cite{Fargion2004},\cite{Fargion2004b}); therefore
   gamma-electron pairs are only partially  regenerated
    by the penetrating muon decay in flight, $\mu^\pm \rightarrow \gamma, e^\pm$
   as a parasite  electromagnetic showering \cite{Cillis2001}.
   Indeed $\mu^\pm $  may decay in flight (let say  at $100$ GeV energy,at $2-3\%$ level within a $12-18$ km distances)
    and they may inject more and more lights, to their primary (far born) shower beam.

   These tens-hundred GeVs  horizontal muons and their associated mini-Cerenkov $\gamma$ Showers have two  main origin:
    (1) either a single muon mostly produced at hundreds km distance by a single (hundreds GeV-TeV
       parental) C.R. hadron : this is a very dominant component;
     (2) a shower by rarer muon, part of a wider and spread horizontal muon  bundle
    of large multiplicity born at TeVs-PeV  or higher energies.
      A whole continuous spectrum  of multiplicity  begins from an unique muon up to a multi muon shower production.
     The  dominant noisy "single" muons at hundred-GeV energies
     will loose memory of the primary low energy and  hidden  mini-shower, (a hundreds GeV or TeVs hadrons );
      a single muon  will blaze just alone.
    The muon "single" rings or arcs frequency is larger (than muon bundles ones) and it is based on solid observational data
    (\cite{Iori04} ; \cite{Grieder01},as shown in fig.2  and references on MUTRON experiment therein); these "noise" event number is:
     $$N_{ev}= \phi_{\mu}(E\simeq 10^{2} eV) \cdot A_{Magic} \cdot \Delta \Omega \cdot
      \Delta(t) \simeq 120 /12 h$$
      The additional gamma  mini-showers around the telescope due to a decay
        (at a probability $p\simeq 0.02$) of those muons in flight, recorded within a
         larger collecting  Area $A_{\gamma} \geq 10^9 cm^2$ is even a more frequent (by a factor $\geq 8$) noisy signal:
       $$N_{ev}\geq \phi_{\mu}(E\simeq 10^{2} eV)\cdot p \cdot A_{\gamma} \cdot \Delta \Omega \cdot
      \Delta(t) \simeq 960 /12 h$$.
        These   single background gamma-showers must take place nearly once
       a minute (in an silent hadronic background) and they are an useful  tool
       to be used as a  meter of the Horizontal C.R. verification.



    On the contrary PeVs (or higher energy) CR shower Cerenkov lights
     may be  observed, more rarely, in coincidence  both by their primary and by their later secondary arc and gamma mini-shower.
   Their $30-100$ GeV  energetic muons are flying  nearly undeflected
  $\Delta \theta \leq 1.6^o \cdot \frac{100 \cdot GeV}{E_{\mu}}\frac{d}{300 km}$
  for a characteristic horizons distances d , partially bent by  geo-magnetic $0.3$  Gauss fields;
  as mentioned, to flight   through the whole horizontal air column depth
  ($360$ km equivalent to $360$ water depth) the muon
   lose nearly $100$ GeV; consequently the origination muon energy should be a little  above this threshold
   to be observed by Magic: (at least $ 130-150 $ GeV along most of the flights).
   The deflection angle is therefore a small one:
    ($\Delta \theta \leq 1^o \cdot \frac{150 \cdot GeV}{E_{\mu}}\frac{d}{300
   km}$). Magic telescope area ($A = 2.5 \cdot 10^6 cm^2$) may record at first approximation the
   following event number of  direct muon hitting the Telescope, flashing  as rings and arcs, each night:
  $$N_{ev}=\phi_{CR}(E= 6\cdot 10^{15} eV)\cdot N_\mu(10^2\cdot GeV) \cdot A_{Magic} \cdot \Delta \Omega \cdot
      \Delta(t) \simeq 45 /12 h$$ to be correlated (at $11\%$ probability) with the above
      results of $401$ primary Cerenkov flashes at the far distances.
   As already mentioned before, in addition the same muons are decaying in flight  at a minimal probability $2\%$
   leading to a  mini-gamma-shower event number in a quite wider  area ($A_{\gamma}= 10^9 cm^2$):
   $$N_{ev}= \phi_{CR}(E= 6\cdot 10^{15} eV)\cdot N_\mu(10^2\cdot GeV) \cdot p \cdot A_{\gamma} \cdot \Delta \Omega \cdot
      \Delta(t) \simeq 360 /12 h$$
      Therefore , in conclusion,  at $87^o-88^o$ zenith angle, there are a flow of
      primary $ E_{C.R}\simeq 6\cdot 10^{16} eV$ C.R. whose earliest
      showers and consequent secondary muon-arcs as well as
      nearby muon-electron mini-shower take place at comparable (one every
      $120$ s.) rate. These $certain$  clustered signals offer an unique tool
      for immediate    gauging and calibrating of Magic (as well as
      Hess,Cangaroo,Veritas Cerenkov Telescope Arrays) for Horizontal High Energy Cosmic Ray Showers.
      Some more rare event may contain at once both Rings,Arcs and tail
      of gamma  shower and Cerenkov of far primary shower.
       It is possible to estimate also the observable muons-electron-Cerenkov  photons
  from up-going  Albedo muons observed by recent ground experiments  \cite{NEVOD}  \cite{Decor}: their flux
   is already suppressed at zenith angle $91^o$ by at least two order of
   magnitude and by four order for up-going zenith  angles $94^o$.
   Pairs or bundles are nevertheless more rare (up to $\phi_{\mu} \leq 3 \cdot 10^{-13} cm^{-2}s^{-1}sr^{-1}$
   \cite{NEVOD}  \cite{Decor}). They are never associated to up-going shower out of the case of
   tau air-showers or by nearby Glashow $\bar{\nu_e}-e\rightarrow W^-$ and  comparable $\chi^o + e\rightarrow \tilde{e}$
   detectable by stereoscopic Magic or Hess array telescopes, selecting and evaluating their column depth origination, just discussed below.

\section{UHE $\bar{\nu_e}-e\rightarrow W^-$ and $\chi^o + e\rightarrow \tilde{e}$ resonances versus $\tau$ air-showers }
  The appearance of horizontal UHE
   $\bar{\nu_\tau}$ ${\nu_\tau}\rightarrow \tau$ air-showers (Hortaus or Earth-Skimming neutrinos)
    has been widely studied  \cite{Fargion1999},\cite{Fargion 2002a},
  \cite{Bertou2002},\cite{Feng2002},\cite{Fargion03},\cite{Fargion2004},\cite{Jones04},
  \cite{Yoshida2004},\cite{Tseng03},\cite{Fargion2004b};  their rise from the Earth is source of rare clear signals for neutrino UHE astronomy (see fig.3).
   However also horizontal  events by UHE $6.3$ PeV, Glashow $\bar{\nu_e}-e\rightarrow W^-$ and a possible
   comparable SUSY  $\chi^o + e\rightarrow \tilde{e}$ \cite{Datta} hitting and showering in air have  non negligible event number:
   $$N_{ev}= \phi_{\bar{\nu_e}} (E= 6\cdot 10^{15} eV) \cdot A\cdot \Delta \Omega \cdot
      \Delta(t) \simeq  5.2 \cdot 10^{-4}/12 h$$
      assuming the minimal  GZK neutrino flux : $\phi_{\bar{\nu_e}} (E= 6\cdot 10^{15} eV)\simeq 5 \cdot
      10^{-15}$ eV $ cm^{-2} s^{-1}sr^{-1}$; the energy flux is $\phi_{\bar{\nu_e}}\cdot E_{\bar{\nu_e}} \simeq $ $30$ eV
$cm^{-2}s^{-1} sr^{-1}$.(We assume  an observing distance at the
horizons $d =167 km \cdot \sqrt{\frac{h}{2.2 km}}$.) Therefore
   during a year of night records and such a minimal GZK flux,  a crown array of
   a $90$ Magic-like telescopes on $2\cdot \pi = 360^o$ circle facing the horizons,
   would discover an event number  comparable to a
   $Km^3$ detector, ( nearly a dozen events a year). Indeed  Magic facing
   at the Horizons as it is, offer a detection comparable to
   present AMANDA $\simeq  1\%  Km^3$ effective volume. In conclusion while Magic looking
   up see down-ward $\gamma$ tens GeVs Astronomy, Magic facing the Horizons may well
   see far  UHE (PeVs-EeVs) CR, and rarely, along the edge, GZK $\bar{\nu_e}-e\rightarrow W^-$
   neutrinos showering in air, as well as charged current
    $\nu_{\tau}+N \rightarrow \tau^-$,$\bar{\nu_{\tau}}+ N \rightarrow \tau^+$ whose decay in flight while in air leads to Hortau  air-showers. Even  SUSY
$\chi^o + e\rightarrow \tilde{e} \rightarrow\chi^o + e $  lights
in the sky (with showers) may blaze on  the far frontier of the
Earth.

\begin{figure}
\begin{center}
\epsfig{figure=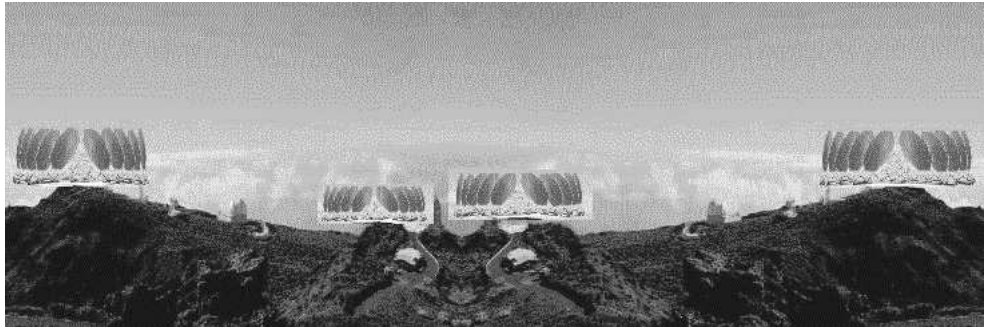,width=13cm} \caption{ An ideal
Array of Crown-Telescopes at Magic site, able to trace the
horizontal showers at different angle of view , few km distance;
these elements may test the contemporaneous shower profile and
they might cooperate with similar scintillator (Crown detectors
not shown in figure, see Iori et all. and  Fargion et all,ApJ.
2004) able to better trigger and reveal the electromagnetic and
the muon content of the showers.
}
   \label{fig5}
   \end{center}
\end{figure}

\end{document}